\documentclass[english]{article}
\usepackage[T1]{fontenc}
\usepackage[latin9]{inputenc}
\usepackage{float}
\usepackage{graphicx}

\makeatletter
\newcommand{\lyxaddress}[1]{
	\par {\raggedright #1
	\vspace{1.4em}
	\noindent\par}
}

\usepackage{babel}

\makeatother

\usepackage{babel}
\begin{document}
\title{A site-site interaction two-dimensional model with water like structural
properties}
\author{Tangi Baré$^{1}$and Maxime Besserve$^{1}$, Tomaz Urbic$^{2}$\thanks{tomaz.urbic@fkkt.uni-lj.si }
and Aurélien Perera$^{1}$ \thanks{Corresponding author:aurelien.perera@sorbonne-universite.fr}}
\maketitle

\lyxaddress{$^{1}$Laboratoire de Physique Théorique de la Matière Condensée
(UMR CNRS 7600), Sorbonne Université, 4 Place Jussieu, F75252, Paris
cedex 05, France.}

\lyxaddress{$^{2}$Faculty of Chemistry and Chemical Technology, University of
Ljubljana, Vecna pot 113, 1000 Ljubljana, Slovenia.}
\begin{abstract}
A site-site interaction model is proposed for water in two-dimension,
as an alternative to the traditional Mercedes-Benz model. In MB model,
water molecules are modeled as 2-dimensional Lennard-Jones disks with
three hydrogen bonding arms arranged symmetrically, resembling the
Mercedes-Benz logo. The MB model qualitatively predicts both the anomalous
properties of pure water and the anomalous solvation thermodynamics
of non-polar molecules. One of the features of this earlier model
was to have a pair correlation function with first peak for the Lennard-Jones
contact distinct of that corresponding to the hydrogen bonding, which
is very different from real water which has a single first peak, but
a dual peak for the structure factor. The site-site model proposed
here reproduces this typical feature of real water, both in real and
reciprocal space. It also reproduces several of the known anomalies
of real water, such as the density maximum. In addition, because of
the screened Coulomb interaction between the sites, the new model
appear to exhibit more homogeneity that the MB models and their variants,
the latter which is highlighted by a $k=0$ increase of their structure
factors. The new model transfers the usual bond order paradigm into
a charge order paradigm, enforcing atom-atom interactions over orientational
interactions. 
\end{abstract}

\section{Introduction}

Liquid water, the most abundant liquid and common liquid on Earth,
is also the most elusive as far as its numerous anomalous properties
are concerned - more than 60 in the sole liquid phase\cite{Ball,Chaplin}.
These anomalous properties are believed to be related to the hydrogen
bonding property and tetrahedrality\cite{DebeNat,MCBF}, which makes
water an associated liquid because of the underlying Hbond ``network''\cite{BernalFowler}.
The hydrogen bonding is crucial to understand the behaviour and properties
of water and aqueous solutions. Despite extensive theoretical efforts
and simulations, how water properties emerge from its molecular structure
remains poorly understood. A large number of models of varying complexity
have been developed and analysed to model water's unusual properties,
for reviews, see \cite{Dill-Review,water}. The key goal of liquid--state
statistical thermodynamics is to develop a quantitative theories for
water and aqueous solutions. It is generally admitted that theory
and simulations have only partly explained how water's molecular structure
leads its density, compressibility, expansion coefficient and heat
capacity as functions of temperature and pressure, including its well
known anomalies. There have been two main approaches to modeling properties
of liquids. One approach is to perform computer simulations of atomically
detailed models. These models aim for realistic detail and include
variables describing van der Waals and Coulomb interactions, hydrogen
bonding, etc. (reviewed in \cite{guillot}). Such approaches can depend
critically on the force--field used in the calculation\cite{millot}.
However, many properties of water and aqueous solutions can also be
captured by simpler models. While the existence and properties of
hydrogen bond network have proven quite difficult to characterise
both experimentally and theoretically, water thermo-physical properties
and structure are usually discussed in terms of the local Hbond and
tetrahedral molecular configurations\cite{StillingerWeber,Stanley}.
Related studies has led to many speculations on the possible underlying
structure of this liquid, both for pure water\cite{Patey} and aqueous
mixtures\cite{bookAqMix}. While it is quite difficult to picture
three dimension water network, the two-dimension Mercedes-Benz (MB)
water model, originally proposed by Ben-Naim in 1971\cite{ben,ben2},
has considerably helped improve this situation, particularly through
the numerous studies by Ken Dill and collaborators\cite{mbs,Tomaz1,Dill2,Tomaz2}.
These studies has helped consolidating intuitive pictures of water
structure and behaviour near hydrophilic and hydrophobic solutes,
as described in the review article\cite{Dill-Review}. The MB model
serves as one of the simplest models of an orientationally dependent
liquid, so it can serve as a test bed for developing analytical theories.
Another important advantage of the MB model, compared to more realistic
water models, is that the underlying physical principles can be more
readily explored and visualized in two dimensions. The MB model was
also extensively studied by analytical methods such as thermodynamic
perturbation theory and integral equation theory \cite{Tomaz1,urbic1,urbic3,urbic4,urbic5,urbic6,urbic7}.

Despite these advances, this model has some disadvantages. The MB
water model is entirely based on the angular orientation of the 3
hydrogen bonding arms, and because of this it requires supplementary
interaction hypothesis in order to be extended to study aqueous mixtures\cite{aqsol2D,aqprot2D},
and in particular electrolytes\cite{2Delectrol1,2Delectrol2}. In
addition, it does not have a pair distribution function $g(r)$ which
looks like that of real water. Indeed, the pair distribution function
of water has very peculiar properties\cite{water-3PS} unlike other
simple molecular liquids. In particular, the first peak is very narrow
and positioned at quite small distance of $r\approx2.8\mathring{A}$,
when the water diameter is $\sigma_{w}\approx3.1\mathring{A}$. This
is a direct consequence of the directionality and strength of the
hydrogen bonding. The $g(r)$ of the MB model has 2 peaks instead
of one, a first narrow peak which corresponds to hydrogen bonding
of the real water, but also a second smaller pre-peak, which corresponds
to the water-water direct contact $\sigma_{w}$. This feature is a
direct consequence of the choice of setting the Hbonding distance
$d_{\mbox{HB}}$ to a value larger than $\sigma_{w}$. This choice
insures that the 2D MB model has a density maximum at low temperature\cite{2D-anomalies},
in analogy with real water.

In addition to these issues, the 2D representation of water reflects
another problem posed by the description of realistic molecular liquids:
the orientational versus the atomic representation. A molecule can
be represented by it orientation through the set of Euler angles $\vec{\Omega},$or
as a set of atoms covalently bonded. This problem is highlighted by
the fact that both descriptions are rigorously equivalent at the level
of the pair interactions: 
\begin{equation}
u(\vec{r},\vec{\Omega}_{1},\vec{\Omega}_{2})=\sum_{i,j}u_{ij}(r_{ij})\label{eq:u}
\end{equation}
where on the lhs the pair interaction is represented in terms of the
intermolecular vector $\vec{r}$ between the 2 molecules, and their
respective orientations through the set of Euler angles $\vec{\Omega}_{1}$
and $\vec{\Omega}_{2}$, and on the rhs the atomic description with
pairs of atoms $i$ and $j$. However, at the level of pair correlations,
the two descriptions are not equivalent. The molecular pair correlation
function $g(\vec{r},\vec{\Omega}_{1},\vec{\Omega}_{2})$ cannot be
expressed in terms of the set of atom-atom pair correlation functions
$g_{ij}(r)$, and each type of function provides a different description
of the molecular liquid properties. This is the difference between
the dipolar or charge-charge description of polar molecular liquids.
Interestingly, the 2D MB water model is purely orientational, and
has no underlying charged site representation, since it was intentionally
built to capture the sole orientational ordering imposed by the hydrogen
bonding in real water. We note that the Rose water model introduced
subsequently\cite{rose0,roseTomaz} is also an orientational model.

In this context, it is highly desirable to have a site-site equivalent
of the MB model, which would allow to introduce other site-site solutes,
which would allow a natural site based interaction model, such as
in the case of ion-water interaction in a 2D electrolyte representation\cite{2Delectrol1,2Delectrol2}.

In this report, we present a site-site interaction model of water,
the SSMB model for site-site Mercedes-Benz model, which is based on
a 7 ``atoms'' representation, with partial charge representation
and screened Coulomb interactions, and which provides a convincing
alternative to the MB model, while showing a closer structural analogy
with real water, in particular in what concerns the $g(r)$. In addition,
the site-site representation allows for a straightforward extension
of three-dimension solute models in terms of atomic representation,
hence permitting to extend and improve the pictorial and intuitive
ideas about water and aqueous mixtures.

The principal focus of this paper is the structural features, and
not so much the thermodynamical and dynamical properties, which, by
construction of the model in analogy with the MB model, should be
preserved more or less similarly, albeit for different state parameters
such as the temperature and the pressure. With this in mind, the remaining
of the paper is as follows. In the next section, the site model is
introduced with a brief reminder of the MB model. Section 3 concerns
details of the simulations. In Section 4, we first remind the structural
properties of the original and modified MB models, before comparing
with the structural properties of the site model. In particular, the
structure factors are shown, which allows to account for the Fourier
representation of the water structure. The final section 5 contains
discussions and our conclusions.

\section{The site-site model}

We first remind the original MB model, as well as a variant of it
which we tried in order to merge the Hbond peak and LJ peak into a
single one, just as real 3D water.

It is important to note that the model has no net dipole moment. Indeed,
in two dimension, it is not possible to have both dipolar ordering
and MB-like orientational ordering. However, this is not problematic
for the case of ion-water interactions, since the asymmetry of the
positive and negative charges allow a natural orientational ordering
of the water model around a cation which differs from that around
the anion. In that, there is a ``circular dipolar'' order, instead
of the linear dipole of real water, which arise from the difference
in orientation of the MB branches and the complementary MB branches.

\subsection{The MB model}

\begin{figure}[H]
\centering \includegraphics[scale=0.4]{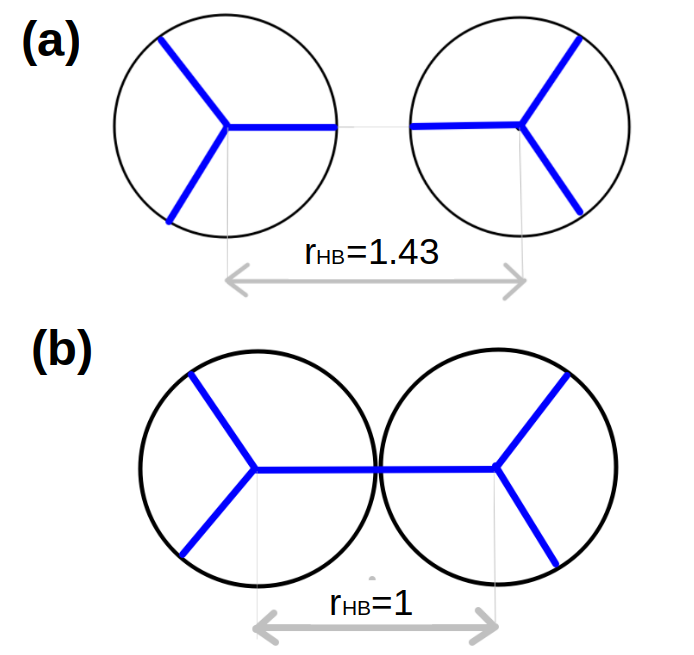} \caption{MB water models in Hbonding configuration: (a) Original MB water model,
and (b) modified mMB model. All distances are expressed in units of
$\sigma$.}

\label{FigMB_mMB} 
\end{figure}

In the original MB model each particle is represented as two--dimensional
disk with Lennard-Jones attraction and three arms arranged as in MB
logo, which can form hydrogen bonds between molecules, as illustrated
in Fig.\ref{FigMB_mMB}. The angle between arms is 120$^{\circ}$\cite{ben,mbs}.
The interaction potential between particles $i$ and $j$ is sum of
a Lennard--Jones (LJ) term and a hydrogen--bonding (HB) term. The
LJ term depends only on the distance between centers of particles
while the HB term also on the orientation of each particle:

\begin{equation}
U_{MB}(\vec{X}_{i},\vec{X}_{j})=U_{\mbox{LJ}}(r_{ij})+U_{\mbox{HB}}(\vec{X}_{i},\vec{X}_{j})\label{uMB}
\end{equation}
where $\overrightarrow{X_{i}}$ is the vector representing the position
and orientation of the $i$-th molecule. The Lennard--Jones part
of the interaction is calculated in a standard way as:

\begin{equation}
U_{\mbox{LJ}}(r_{ij})=4\epsilon\left[\left(\frac{\sigma}{r_{ij}}\right)^{12}-\left(\frac{\sigma}{r_{ij}}\right)^{6}\right]\label{uLJ}
\end{equation}
where $\sigma$ is the diameter of the LJ disc and $\varepsilon$
is the depth of the LJ potential. The hydrogen bond term is sum of
all interactions $U_{HB}^{kl}$ between the arms $k$ and $l$ of
molecules $i$ and $j$, respectively 
\begin{equation}
U_{\mbox{HB}}(\vec{X}_{i},\vec{X}_{j})=\sum_{k,l=1}^{3}U_{\mbox{HB}}^{kl}(r_{ij},\theta_{i},\theta_{j})\label{uMB_ang}
\end{equation}
Gaussian functions are used to model arm--arm interactions, which
depend on orientation of each molecule as well as distance between
molecules 
\begin{equation}
U_{\mbox{HB}}^{kl}(r_{ij},\theta_{i},\theta_{j})=\epsilon_{\mbox{HB}}G(r_{ij}-r_{\mbox{HB}})G(\hat{i}_{k}.\hat{r}_{ij}-1)G(\hat{i}_{l}.\hat{r}_{ij}+1)\label{uMB_ang1}
\end{equation}
where $G(x)$ is an unormalized Gaussian function:

\begin{equation}
G(x)=\exp\left(\frac{-x^{2}}{2\sigma^{2}}\right)\label{G_MB}
\end{equation}
$\varepsilon_{HB}$ is the HB energy and $r_{HB}$ is a HB distance.
$\hat{r}_{ij}$ is the unit vector along $\overrightarrow{r}_{ij}$
and $\overrightarrow{i}_{k}$ is the unit vector representing the
$k$-th arm of the $i$-th particle. Scalar product can be calculated
as 
\[
\hat{i}_{k}.\hat{r}_{ij}=\cos(\theta_{i}+\frac{2\pi}{3}(k-1))
\]
where $\theta_{i}$ is the orientation of $i$-th particle. The strongest
HB between molecules is formed when arms that form bond are parallel
and pointed toward centers of molecules, while distance between centers
of molecules is equal to $r_{HB}$. Throughout this paper, we use
conventional distance and energy parameters which are the diameter
$\sigma$ and the LJ energy $\epsilon_{LJ}$, according to which all
distances are expressed in terms $\sigma$ and energies in terms of
$\epsilon_{LJ}$.

We used two versions of the MB model, one same as original MB model
with the parameter for hydrogen-bond energy, $\varepsilon_{HB}=-10\epsilon_{LJ}$,
and for hydrogen-bond length $r_{\mbox{HB}}=1.43\sigma$ as shown
in Fig.\ref{FigMB_mMB}(a), and the other called mMB in (b) where
use use $\varepsilon_{HB}=-10\epsilon_{LJ}$ and $r_{\mbox{HB}}=\sigma$
.

One of the problems with the original MB model is that the forcing
of optimum Hbonded molecules to be $r_{HB}=1.43\sigma$ apart leads
to a very low density water. Higher densities are possible, but with
strongly diminished possibility of Hbonded configurations. Although
this is not realistic, this original model allows to capture and pictorially
illustrate so many features of the real water, that the low density
was not seen as a problem. A more realist model is to enforce $r_{HB}=\sigma$,
which allows tighter water packing.

\subsection{The SSMB model}

\begin{figure}[H]
\includegraphics[scale=0.3]{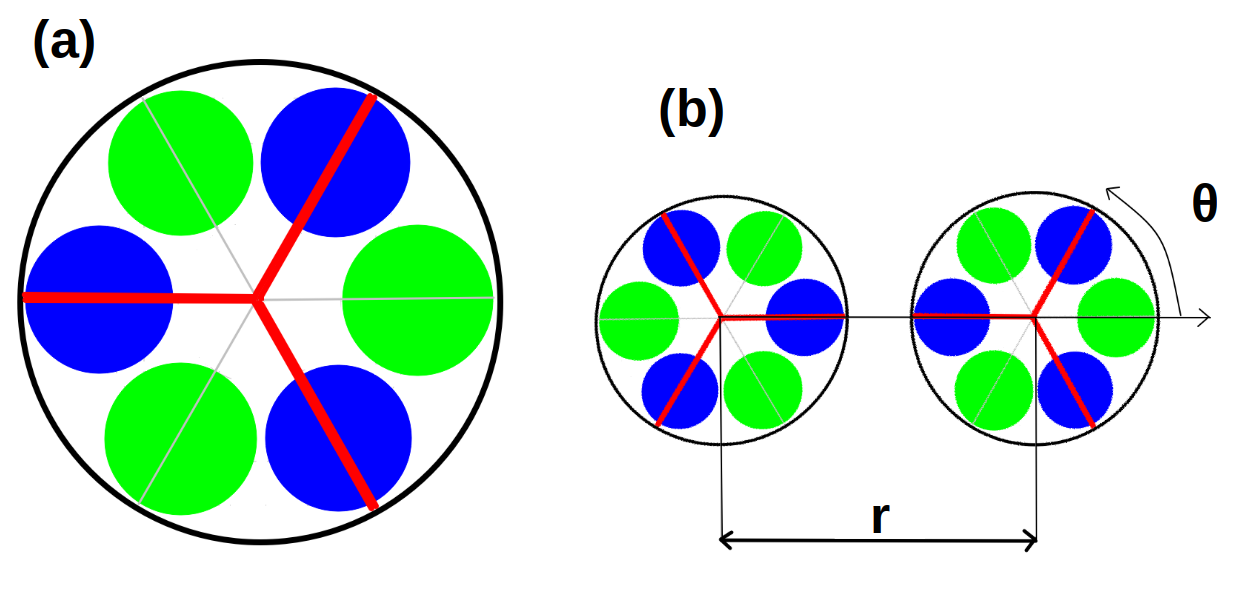}

\centering\caption{The SSMB 1+6 site water model. (a) The red arms correspond to those
of the MB model and are aligned on the blue sites which attract themselves
across different molecules. All other color combinations (green-green
and blue-green) are repulsive. (b) Two interacting SSMB waters with
aligned arms.}

\label{FigSSMB} 
\end{figure}

This model is illustrated in Fig.\ref{FigSSMB}. It is made of 6 small
``charged'' sites and a central LJ site. The small peripheral sites
alternatively bear positive (green) and negative(blue) ``charges''.
These pseudo-charges are used to mimic the original MB branch interactions.
They do not correspond to real Coulomb charges per say. Indeed, in
this SSMB model, blue negative charges attract each other, while all
other charge combinations are Coulomb repulsive. We use a screened
Coulomb (Yukawa) interaction.

The blue charges form the original MB arms (in red in Fig.\ref{FigSSMB}),
thus preserving the same orientation parameters. However, the total
interaction can be now be written as the sum over all pairs of sites:

\begin{equation}
U_{SS}(\vec{X}_{1},\vec{X}_{2})=U_{LJ}(r_{12})+\sum_{i_{1}=1}^{6}\sum_{i_{2}=1}^{6}U_{C}(r_{i_{1}i_{2}})\label{uSS}
\end{equation}
with the ``Coulomb'' part having the Yukawa form with a core 12
repulsion (to avoid charge collapse): 
\begin{equation}
U_{C}(r_{ab})=\epsilon_{ab}\left(\frac{\sigma_{ab}}{r_{ab}}\right)^{12}+\alpha(a,b)\frac{\exp(-r_{ab}/\kappa_{ab})}{r_{ab}}\label{uCoul}
\end{equation}
where $a$ and $b$ are the (blue/green) site index for each molecule
1 and 2, and the various parameters are as follows. The core repulsive
parameters follow the usual Lorentz-Berthelot form, with $\sigma_{ab}=\left(\sigma_{a}+\sigma_{b}\right)/2$
and $\epsilon_{ab}=\sqrt{\epsilon_{a}\epsilon_{b}}$.

The ``Coulomb'' coefficients $\alpha(a,b)$ are defined as

\begin{equation}
\alpha_{ab}=557s_{ab}Z_{a}Z_{b}\label{alpha_ab}
\end{equation}
where $Z_{a}$$(Z_{A}>0)$ is the ``valence'' of site $a$, $s_{ab}$
is the sign of the interaction defined as described above: 
\begin{equation}
\begin{array}{ccc}
s_{blue-blue} & = & -1\\
s_{green-green} & = & +1\\
s_{blue-green} & = & +1
\end{array}\label{sab}
\end{equation}
This way, blue sites always attract each other, while any other combination
lead to a repulsive interaction. The coefficient $557$ represents
the ``Coulomb'' magnitude as defined in our previous works\cite{our2D1,our2D2}.
A model with true Coulomb charges will lead to sites of opposite colors
to attract, unlike in the MB model, whereas a model with pure imaginary
charges $iZ$ would satisfy like site attractions. However, such model
would also imply that green sites would attract each other, thus introducing
a severe bias from the MB model spirit. It is for these reasons that
the additional coefficient $s_{ab}$ needs to introduced.

The model parameters are then the energy $\epsilon_{a}$, diameter
$\sigma_{a},$ Yukawa $\kappa_{a}$ and valence $Z_{a}$ of the blue
and green sites, which is $8$ parameters to adjust in order to recover
an interaction similar to that of the mMB model.

All the the charged sites are positioned at $d=\sigma/3$ from the
center of the molecule. The ($x$,$y$) coordinates of the blue sites
are ($d$,$0$), ($-d/2$,$d\sqrt{3}/2$) and ($-d/2$,$-d\sqrt{3}/2$),
while the coordinates of the green sites are ($-d$,$0$), ($d/2$,$d\sqrt{3}/2$))
and ($d/2$,$-d\sqrt{3}/2$).

We have selected the parameters as to fit to the closest the mMB model
energies for all orientations and distances of 2 water molecules.
The following parameters have been used: $\epsilon_{a}=\epsilon_{LJ}/2$,
$\sigma_{a}=\sigma/3$, $\kappa_{a}=0.15\sigma$, $Z_{blue}=0.30$,
$Z_{green}=0.115$. 

\subsection{Comparison between the MB, mMB and SSMB models}

\begin{figure}[H]
\includegraphics[scale=0.4]{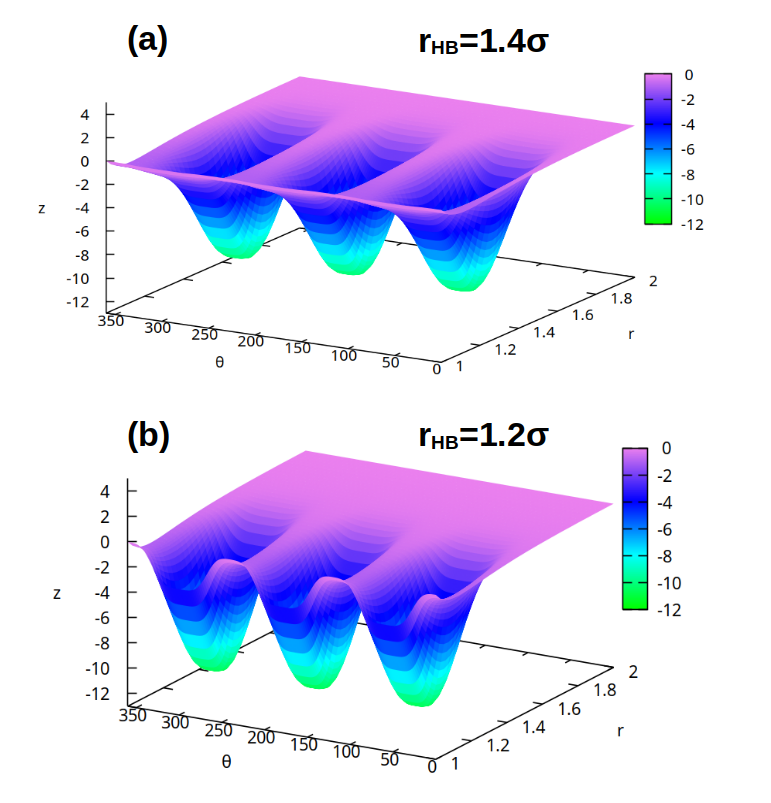}

\caption{3-dimension representations of the original MB interaction $U_{MB}(r,\theta_{1}=0,\theta)$
where the water molecule is at fixed orientation $\theta_{1}=0$ ,
while both the intermolecular distance $r$ and the rotation $\theta_{2}=\theta$
of the second molecule are varied. (a) with the original parameter
$r_{HB}=1.4\sigma$, and (b) with the smaller value $r_{HB}=1.2\sigma$,
hence showing the influence of this parameter.}

\label{Fig.potMB} 
\end{figure}

Fig.\ref{Fig.potMB} shows a comparison of the interaction \ref{uSS}
for the original MB model for 2 different values of $r_{HB}$, in
a 3-dimensional representation, for fixed orientation of the first
water molecule, when the orientation $\theta$ of the second molecule
and its distance $r$ to the first one, are both varied. This representation
highlights the shifted hydrogen bonding wells for the orientations
$\theta=60$, $120$ and $240$, away from then core contact. This
shift is thought as an extension of that of the LJ attractive well,
positioned at some distance from the core contact.

\begin{figure}[H]
\includegraphics[scale=0.4]{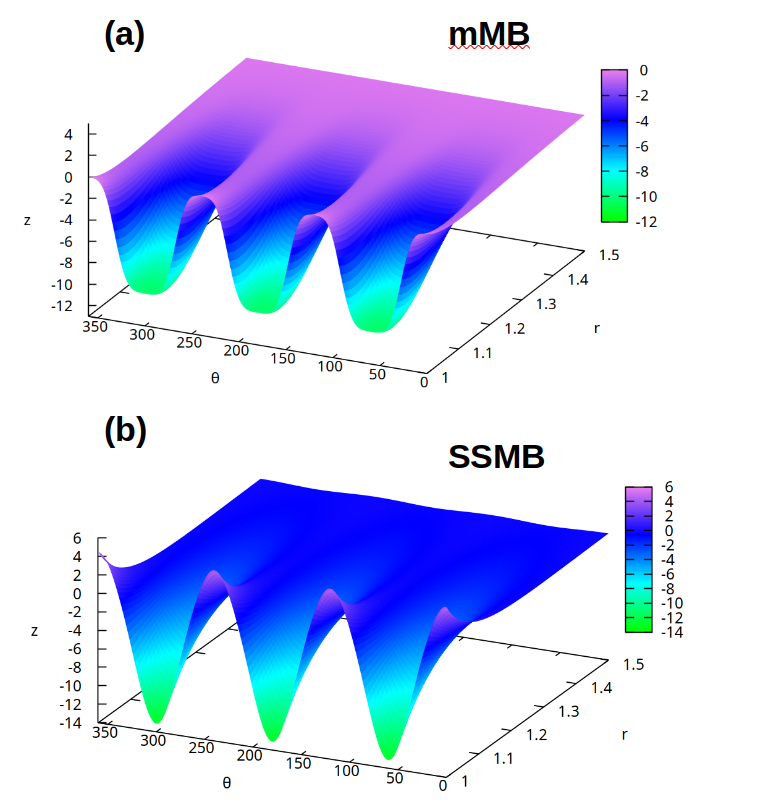}

\caption{Comparison of the 3-dimension representations of the modified mMB
interaction model (a), with $r_{HB}=1.0\sigma$ and the site-site
model (b). The conventions are as in Fig.\ref{Fig.potMB}.}

\label{Fig.potMBcomp1} 
\end{figure}

Fig.\ref{Fig.potMBcomp1} shows a similar 3-dimension comparison between
the modified mMB model (with $r_{HB}=\sigma$) and the present site-site
MB model. The attractive wells are now moved closer to the core $\sigma=1$,
as in the real 3D water models. We note that the Gaussian well are
more ``rounded'' than the Yukawa ones, while this second model shows
more repulsion between the MB arms.

\begin{figure}[H]
\includegraphics[scale=0.4]{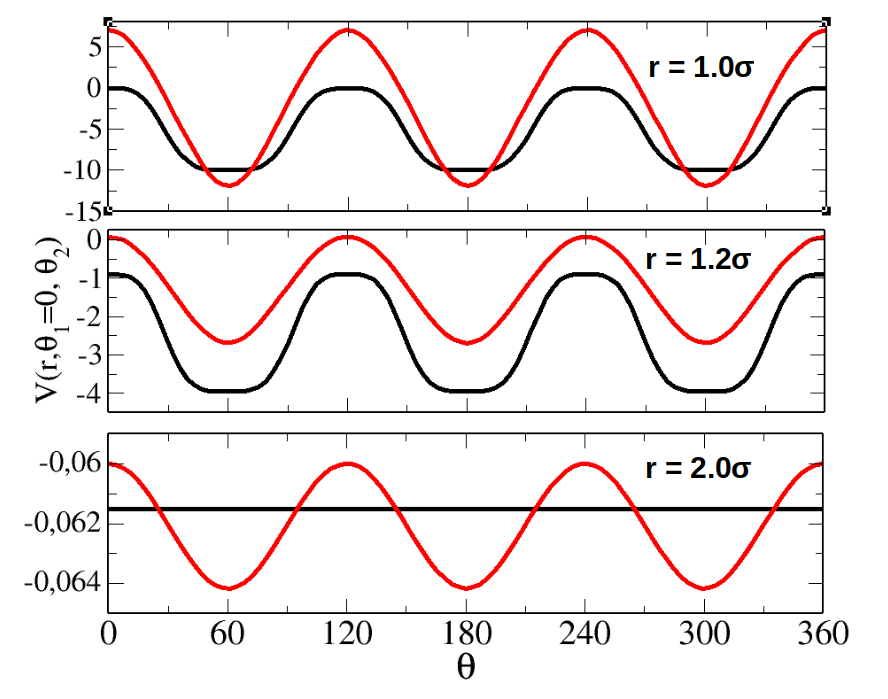}

\caption{Comparison of the mMB model (black lines) with the SSMB model (red
lines), for 3 different intermolecular distances: (top) $r=1.0\sigma$,
(middle) $r=1.2\sigma$ and (bottom) $r=2\sigma$.}

\label{Fig.potMBcomp2} 
\end{figure}

This difference is highlighted in Fig.\ref{Fig.potMBcomp2}, which
shows cuts the previous 3D plots, for specific intermolecular distances
$r=\sigma$, $1.2\sigma$ and $2\sigma$. The SSMB model is slightly
more attractive at the Hbonding angles $\theta=60,120$ and $240$,
and more repulsive than the mMB model between the arms, but also at
short distance $1.2\sigma$. Since the Gaussian interaction decays
faster, one notice the longer range screened-Coulomb type interaction
of the Yukawa interaction in the lower panel of Fig.\ref{Fig.potMBcomp2}.

This long range behaviour should be consistent with the Coulomb nature,
when considering other types of molecules, such as ionic species for
example, such as those we have considered in our previous works\cite{our2D1,our2D2}.

\section{Monte Carlo simulation and technical details}

Monte Carlo (MC) simulations with Metropolis algorithm were performed
in order to determine thermodynamical properties of the MB model in
the isothermal-isobaric (N,p,T) ensemble. Pair correlation functions
were computed in the Canonical (N,V,T) ensemble. Periodic boundary
conditions and minimum image convention were used to mimic macroscopic
system. For NPT ensemble simulations, system sizes from $N=100$ up
to $N=400$ water particles were used. This number of particles in
2D is equivalent to $1000$-$8000$ particles in 3D. For NVT ensemble
simulations, it was needed to go up to $N=441$ in order to avoid
artifacts in the calculated structure factors. The initial positions
of particles were randomly chosen in a way where there was no overlap
between molecules. A molecule was randomly chosen in each MC step
in order to be translated or rotated. On average, each cycle consisted
of one rotational and translational attempt per particle and one attempt
to change volume of system. First, the system was allowed to equilibrate
for minimum of $3\times10^{4}$ - $10\times10^{4}$ cycles, which
depended on the temperature of the system. After the system was equilibrated,
the sampling was performed in 10-20 series, each consisting of minimum
$3\times10^{4}$ cycles. Mechanical properties such as enthalpy and
volume were calculated as the statistical averages of these quantities
over the course of the simulations \cite{hansen,frenkel}. Heat capacity,
$C_{p}$, isothermal compressibility, $\kappa_{T}$, and thermal expansion
coefficient, $\alpha_{T}$ are computed from the fluctuation formulas
of enthalpy, $H$, and volume, $V$ \cite{mbs}.

\begin{equation}
\begin{array}{l}
c_{p}=\frac{<H^{2}>-<H>^{2}}{NT^{2}}\\
\\
\kappa_{T}=\quad\frac{<V^{2}>-<V>^{2}}{T<V>}\\
\\
\alpha_{T}=\quad\frac{<VH>-<V><H>}{T^{2}<V>}
\end{array}\label{eq:fluctuations}
\end{equation}
The structure factors $S(k)$ were calculated through the usual Talman
transform \cite{Talman,2DElli} of the pair correlation function $g(r)$.
Since this transform necessitates that the distances are sampled in
logarithmic spacing, the regularly spaced $g(r)$ obtained from the
computer simulations are interpolated on the logarithmic scale. As
a test of this accuracy of the interpolation, we have checked that
the $S(k)$ for a simple LJ liquid would be consistent with that obtained
from integral equation theories such as Percus-Yevick or Hypernetted-chain
equations. This technique was used in our previous works in \cite{our2D1,our2D2}.
Small oscillations often appear in the low-k limit, which arise partly
from the Talman technique itself \cite{Talman}, but also from either
truncated oscillatory behaviour of $g(r)$ in small boxes. The only
remedy is then to simulate larger systems, such that these packing
structural oscillations become acceptably small.

\section{Results}

We first recall the structural features of the original MB model,
as well as the modified MB model. The reduced density $\rho$ is defined
a for the LJ system by $\rho=(N/S)\sigma^{2},$where N is the number
of particles per surface S. This definition differs from that used
in the original and subsequent paper, where it is defined with respect
to the Hbonding distance $r_{HB}$ as $\rho_{MB}=(N/S)r_{HB}^{2}$
. Similarly, the reduced temperature T is defined in terms of the
LJ temperature $T_{LJ}=\epsilon_{LJ}/k_{B}$, where $\epsilon_{LJ}$
is the LJ energy as in Eq.(\ref{uLJ}) and $k_{B}$ is the Boltzmann
constant. The temperature scale is set by $T_{LJ}=1$. This is again
different from the original papers where the temperature is scaled
by $T_{MB}=\epsilon_{MB}/k_{B}$, where $\epsilon_{MB}$ is the Hbonding
energy in Eq.(\ref{uMB_ang}). 

To summarize, the following scaling correspondence are required to
compare the present units with that of previous MB papers: $T=10T_{MB}$,
$\rho=0.49\rho_{MB}$, and $r=0.7r_{MB}$.

\subsection{Structure of the original MB model}

\begin{figure}[H]
\includegraphics[scale=0.4]{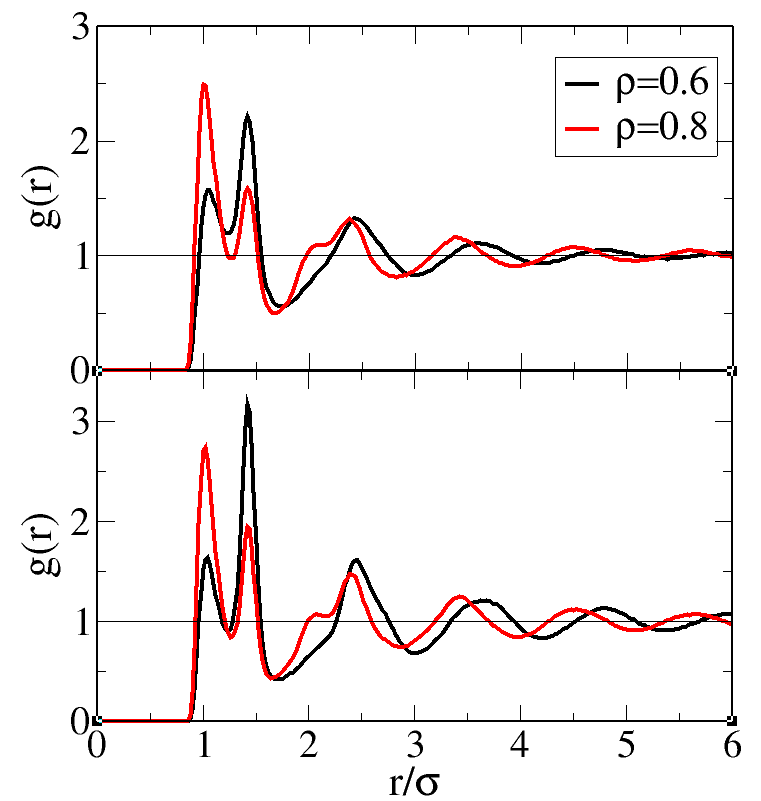}

\caption{Pair correlation function of the MB water model for 2 different densities
and two temperatures $T=3$ (top panel) and $T=2$ (lower panel).
Corresponding data was obtained with $N=441$ particles to ensurre
proper statistics.}

\label{Fig_grMB} 
\end{figure}

As illustrated in Fig.\ref{FigMB_mMB}, the original MB model with
$r_{HB}/\sigma=1$ imposes 2 distinct contact distance, one which
is the Hbonding distance at $r=r_{HB}$, and the other the disc-to-disc
contact at $r=\sigma$. This is illustrated in Fig.\ref{Fig_grMB}
through the pair correlation functions for this model, for 2 different
densities and 2 different temperatures. The density $\rho=0.8$ correspond
to a typical dense liquid, while $\rho=0.6$ is more fluid-like density.
In the original MB model, the particles are spaced quite apart because
of the large Hbonding distance $r_{HB}=1.4\sigma$. The dual contact
distance is quite apparent through the split first peak in the g(r),
when the first peak corresponds to the distance $r/\sigma\approx1$,
while the second to $r/\sigma\approx1.4$. For low density (black
curves), it is the Hbond peak which is increased since particles can
stay far apart. However, when reaching high packings (red curves),
the large Hbonding distance is harder to achieve and it is the first
peak which increase. This inversion of the peak increases when reducing
the temperature.

\begin{figure}[H]
\includegraphics[scale=0.4]{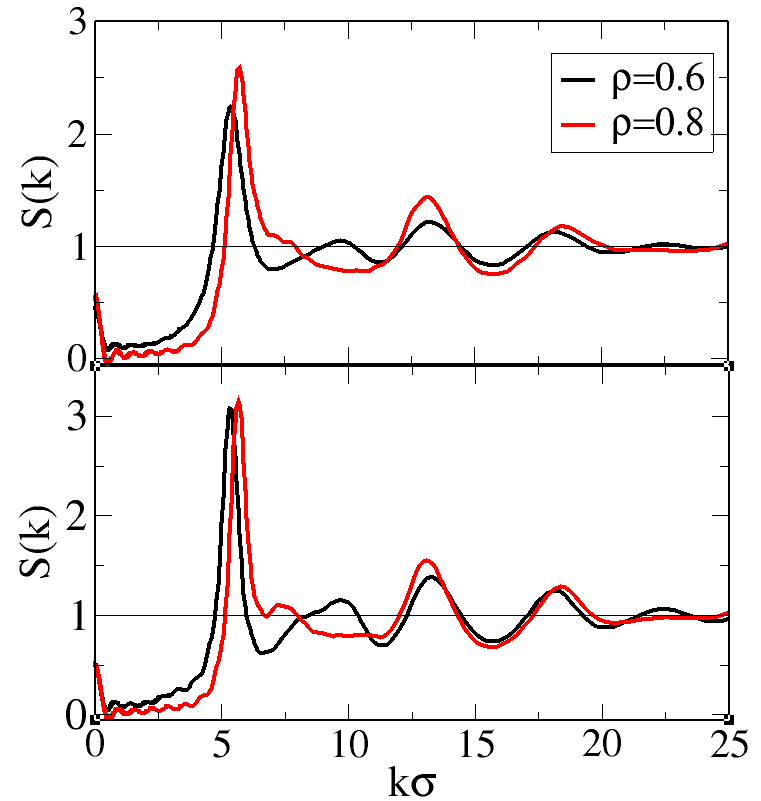}

\caption{Structure factor of the MB water model for the same state parameters
and conventions as in Fig.\ref{Fig_grMB}}

\label{FigSkMB} 
\end{figure}

The corresponding structure factors are shown in Fig.\ref{FigSkMB}.
We note that the large $k$ oscillations are not regular, which arise
from the dual peak feature in the $g(r)$. We equally note the raise
at $k=0$. Such a raise is usually intepreted in terms of the existence
of large density fluctuations. The presence of bonding interactions
can also appear as a density fluctuation, since it would increase
local heterogeneity. This feature which is also found in real water,
but only at low temperatures\cite{small-k-1,small-k-2}. It would
seem that the MB model over emphasizes such cluster-fluctuations in
a wider range of densites and temperatures. We attribute this to the
interaction induced heterogeneity in Section 4.5.

Another important feature of this model is the main peak of the structure
factor, which do not reflect the dual bonding correlations, as it
does for the real water.

We note that small spurious oscillations, particularly for the low
temperature case, which are more pronounced for the lower temperatures
and higher densities. Similar artifacts are due to 2D Fourier transform
techniques, as discussed in Section 3, and are usually too small to
be noticed. However, in the present case, the $k=0$ raise, which
is a genuine physical effect due to fluctuations and clustering, amplifies
these oscillations to the point they become noticeable. It was necessary
to compute $g(r)$ for a larger system of $N=441$ particles, in order
to minimize these artifacts, which is an indication of the importance
of properly sampling fluctuations for the MB model. These artifacts
do not hinder our comprehension of the structure in this model.

\subsection{Structure of the modified MB model}

\begin{figure}[H]
\includegraphics[scale=0.4]{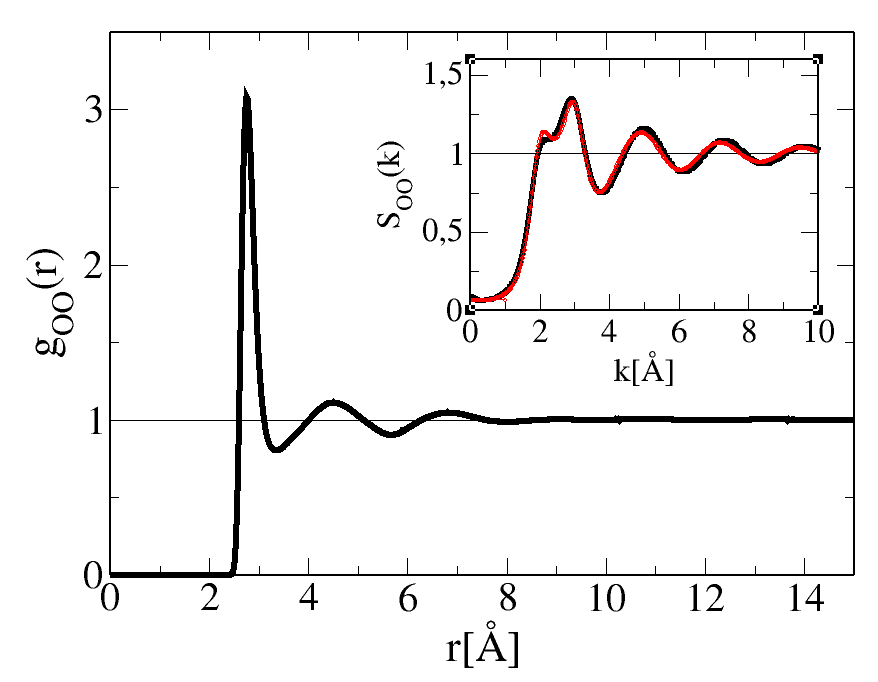}

\caption{Structure functions of the SPC/E water model. The red curve shows
the experimental x-ray structure factor from Ref\cite{x-ray_water}.}

\label{FigSPCE} 
\end{figure}

The appearance of the dual peak in $g(r)$ for the MB model and its
absence in the $S(k)$ main peak are typical feature of this model,
and are at variance with that of real water, as illustrated in Fig.\ref{FigSPCE},
which shows the oxygen-oxygen pair correlation function and structure
factor for the SPC/E model, the latter which is in good agreement
with the x-ray experimental data \cite{x-ray_water,guillot}. Both
data are taken at ambient conditions for real water. The SPC/E model
is one of the popular water models, which represent relatively well
most water properties\cite{guillot}. In the real functions, the split
peak is transferred to the main peak of $S(k)$, while the main peak
of $g(r)$ is very narrow but with no splitting. Another important
feature of the $g(r)$, noted in Ref{[}\cite{water-3PS}{]} is the
abrupt flattening of the large $r$ part of $g(r)$ beyond $r>10\mathring{A}$.
This was interpreted as sharp de-correlation of the Hbond beyond $r\approx10\mathring{A}$,
despite the fact that water is quite dense, with $\rho\approx0.9$.

In order to recover this feature, we reconsider the MB model, but
now allow the shorter Hbonding distance of $r_{HB}/\sigma=1$. We
call this modified MB model the mMB model. In this case, the Hbonding
distance is very close to the disc-to-disc contact $\sigma$, just
like in the real water.

\begin{figure}[H]
\includegraphics[scale=0.4]{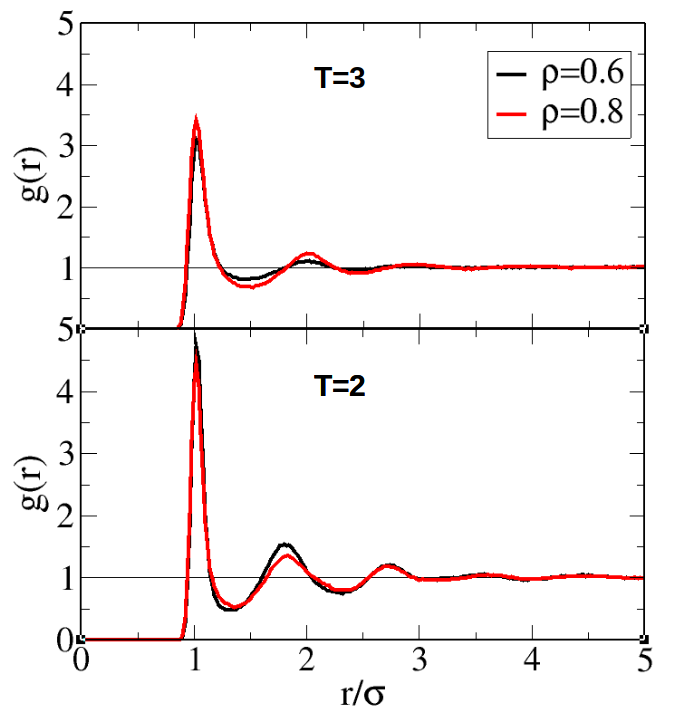}

\caption{Pair correlation functions of the modified mMB model for $r_{H}=1$,
with same state parameters as in Fig.\ref{Fig_grMB}}

\label{FigGrmMB} 
\end{figure}

\begin{figure}[H]
\includegraphics[scale=0.4]{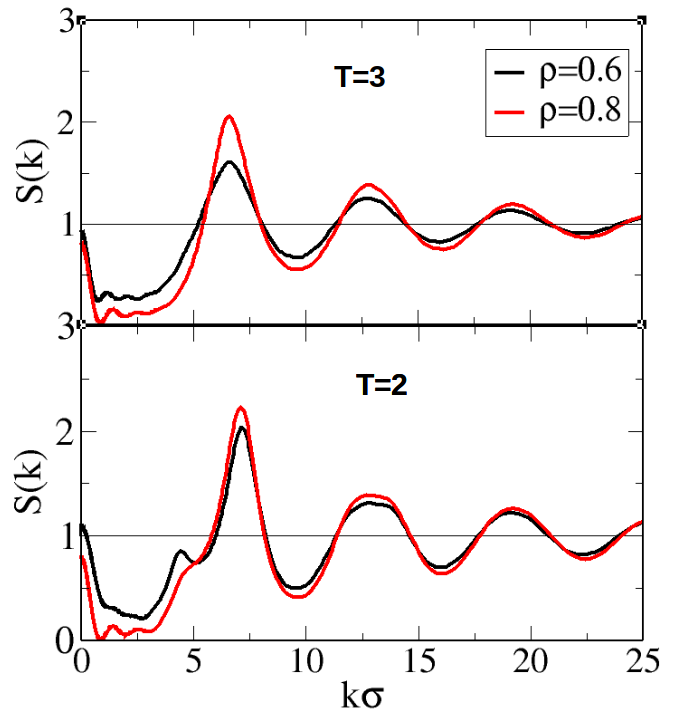}

\caption{Structure factors of the modified mMB model for r\_H=1, with same
state parameters as in Fig.\ref{Fig_grMB} }

\label{FigSkmMB} 
\end{figure}

The pair correlation function for this model are shown in Fig.\ref{FigGrmMB},
while the structure factor are shown in Fig.\ref{FigSkmMB}, for the
same 2 densities and temperature as previously for the MB model. We
note that the $g(r)$ are now strikingly similar to that of the real
water in Fig.\ref{FigSPCE}, with both a sharp first peak and a rapid
loss of correlations beyond $r/\sigma\approx3$. For $\sigma\approx3\mathring{A}$
for real water, this correspond to nearly the same distance of $10\mathring{A}$
in Fig.\ref{FigSPCE}. We also note that there is a much weaker influence
of density and temperature than for the MB model, which indicates
a system structurally dominated by the Hbonding mechanism.

The structure factors in Fig.\ref{FigSkmMB} equally show interesting
features. First of all, for the low temperature $T=2$ case, we note
that the main peak develop a weak outer shoulder, exactly as the real
water in the inset of Fig.\ref{FigSPCE}. This shoulder is absent
at the higher temperature, which shows the destruction of Hbonding
cluster by heating. Next, we see that the large $k$ behaviour is
very regular, which is a direct signature of the packing conditioned
by $\sigma$ instead of a dual distance as in the original MB model.
Finally, we note the existence of the small $k$-raise very similar
to that of the MB model. Since the $k=0$ raise is indicative of density
fluctuations, we see that the Hbond clustering can be interpreted
as a form of density fluctuations. 

The additional small wiggles seen in Fig.\ref{FigSkmMB}, are equally
observed in the present case, and are also amplified by the $k=0$
raise for the same reason as explained before for the MB model.

\subsection{Structure of the site-site MB model}

\begin{figure}[H]
\includegraphics[scale=0.5]{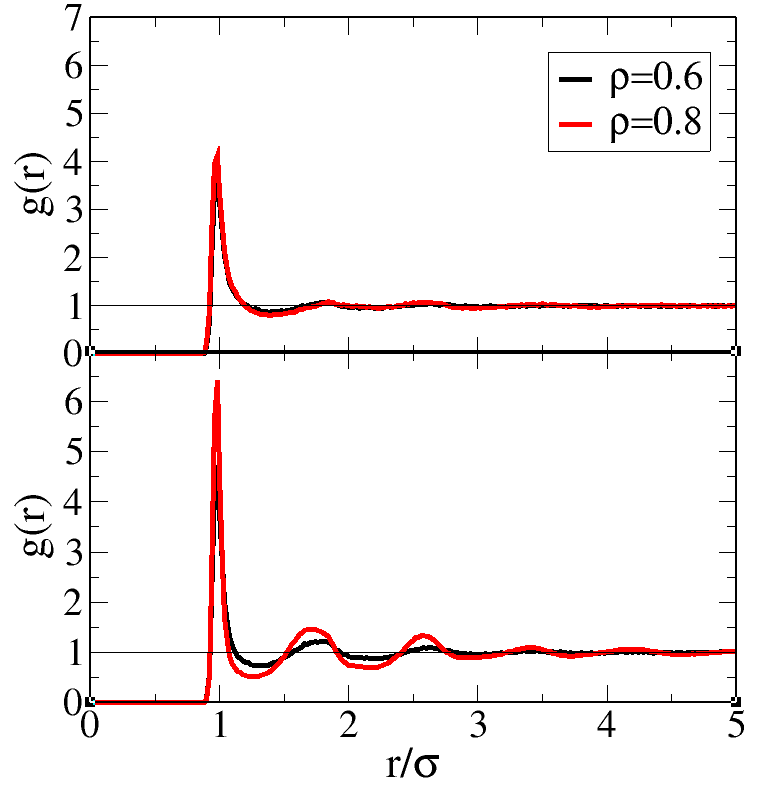}

\caption{Center-to-center pair correlation functions for the site-site SSMB
water model }

\label{FigGrSSMB} 
\end{figure}

\begin{figure}[H]
\includegraphics[scale=0.5]{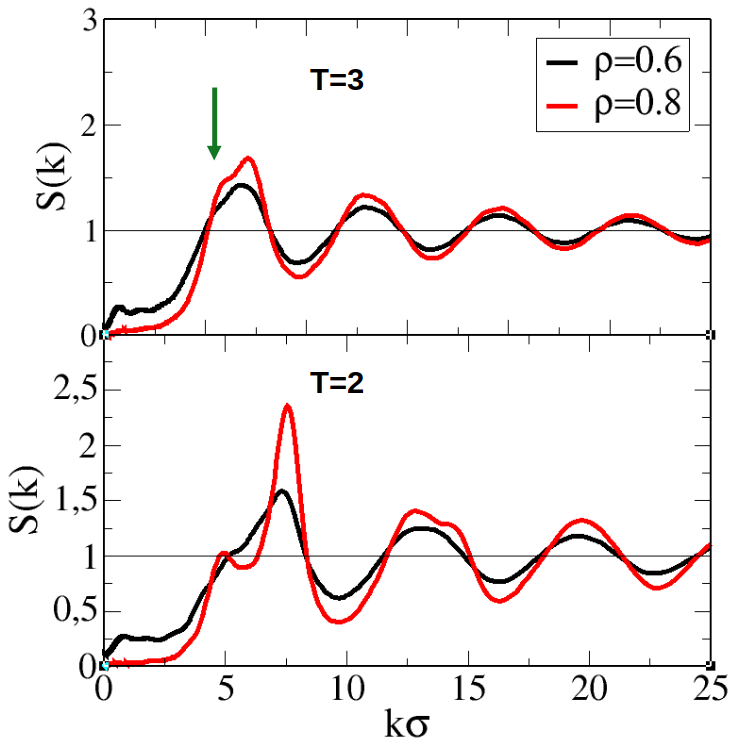}

\caption{Structure factors of the SSMB model for same state parameters as in
Fig.\ref{FigGrSSMB}. The green arrow highlights the shoulder peak
similarity with the structure factor of real water (see inset of Fig.\ref{FigSPCE}).
System size of $N=100$ was used, except for $T=2$ and $\rho=0.8$,
for which $N=221$ was found optimal.}

\label{FigSkSSMB} 
\end{figure}

We now focus on the structural features of the site-site SSMB water
model. This is illustrated in Fig.\ref{FigGrSSMB} for the pair correlation
functions $g(r)$, and in Fig.\ref{FigSkSSMB} for the corresponding
structure factors. Due the similarities with the mMB model imposed
as illustrated in Fig.\ref{Fig.potMBcomp1} and Fig.\ref{Fig.potMBcomp2},
we see that the structure functions are also very similar to that
of the mMB model. We note, however, that the SSMB has features closer
to that the the real water. In particular, we observe a rapid loss
of long range correlations, which is a direct consequence of charge
ordering, as shown in Ref.{[}\cite{water-3PS}{]}.

The similarity with real water is even more striking for the structure
factor $S(k)$ in Fig.\ref{FigSkSSMB}, particularly for $T=3$ and
$\rho=0.8$. Since the real water data corresponds to ambient conditions,
it is tempting to consider these state parameters could equally correspond
to 2D ambient conditions. We observe the same dual first peak, which
witnesses the dual hydrogen bonding and core contact possibilities
for water. At lower temperature, the SSMB model appear more structured
that the mMB model.

We note, however that the $k=0$ behaviour is very different from
that of the MB and mMB models, which both showed a rapid raise, suggesting
strong density fluctuations, even in equivalent ambient conditions.
The SSMB model appears to be closer to the real water, since it shows
the absence of such fluctuations. We also note the strong diminution
of small-$k$ artifacts, which do not get amplified by the $k=0$
raise, absent from this SSMB model.

\subsection{Charge order in the site-site MB model}

\begin{figure}[H]
\includegraphics[scale=0.5]{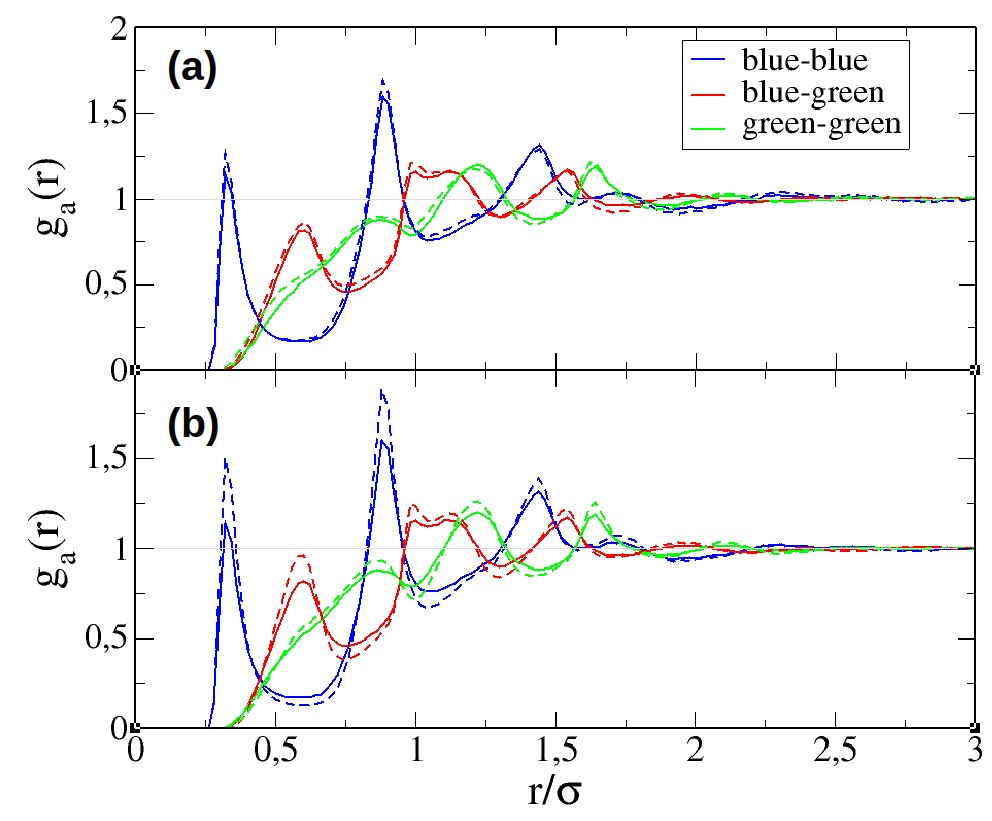}

\caption{Site-site correlation functions. (a) comparison of the density dependence
for $T=3$; full curves for $\rho=0.6$ and dashed curves for $\rho=0.8$.
(b) comparison of the temperature dependence for $\rho=0.$6; full
curves for $T=3$ and dashed for $T=2$. The blue site correlations
are in blue, the green site correlations in green, and cross blue-green
correlations in red.}

\label{FigSItSIt} 
\end{figure}

We now turn toward the ``charged'' site-site correlations, which
should witness site correlations more directly than the center-to-center
correlations we analyzed so far. Fig.\ref{FigSItSIt} shows a comparison
of the density dependence in the upper panel for $T=3$, and in the
lower panel a temperature dependence for $\rho=0.6$. It is found
that the density dependence of these functions is smaller than their
 temperature dependence. It corroborates with the intuitive idea that
Hbond ordering dominates the particles packing, hence reducing the
density dependence, while temperature has a stronger influence in
the formation and destruction of the Hbonds.

The positions and magnitudes of the various peaks is interesting to
analyze. The blue curves have 2 principal peaks, the first one at
$r=\sigma/3$, which is the expected blue-blue contact, and the second
one at $r\approx\sqrt{(}7)/3$, which is the other blue-blue distance
for a typical Hbonding configuration in Fig.\ref{FigSSMB}b. Although
this second peak is higher than the first one, it represents smaller
correlations, since it counts for the 2 possible distances. All other
peaks obey similar criteria. These correlation functions provide a
strong underlying support for the alignment of the arms supposed to
mimic that of the MB model.

Fig.\ref{FigSItSIt} equally shows an important property of the SSMB
model, which is charge ordering. Indeed, it can be observed that the
blue and red curves are in phase opposition, hence witnessing an alternate
spatial disposition of the blue and green sites. This is similar to
what is seen in the case of charge order in real ionic liquids, both
in 3D\cite{myIonic2} and 2D \cite{our2D1,our2D2}. It is interesting
that the SSMB model should demonstrate this type of ordering since
it does not contain true charges. This is particularly interesting
for the perspective of modeling 2D electrolytes with the SSMB water
model.

\begin{figure}[H]
\includegraphics[scale=0.5]{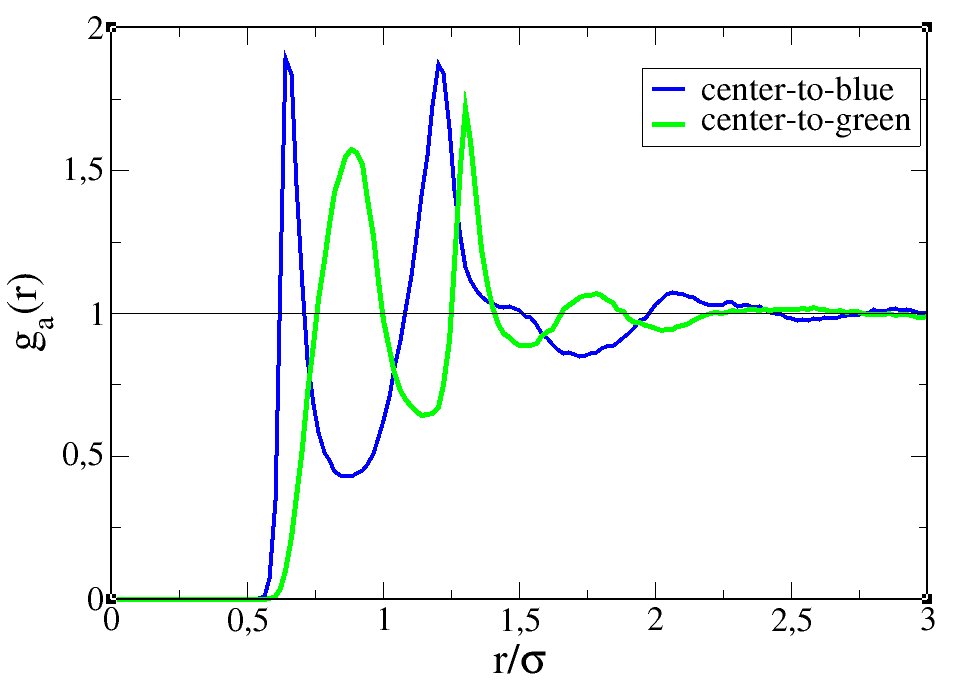}

\caption{Center to site correlation functions for $T=3$ and $\rho=0.6$, providing
another signature of charge ordering (see text).}

\label{FigCenSit} 
\end{figure}

Fig.\ref{FigCenSit} shows the correlation functions from the molecular
center to the auxiliary sites, for the case of $T=3$ and $\rho=0.6$.
We observe again that the blue and green site correlations are in
phase opposition, which is expected as a consequence of phase opposition
between the blue and green sites witnessed in Fig.\ref{FigSItSIt}.
The narrower blue peaks witness the strong blue-blue alignment, while
the broader green peak witness the more dispersed center to green
sites correlations.

\subsection{Snapshots}

\begin{figure}[H]
\includegraphics[scale=0.3]{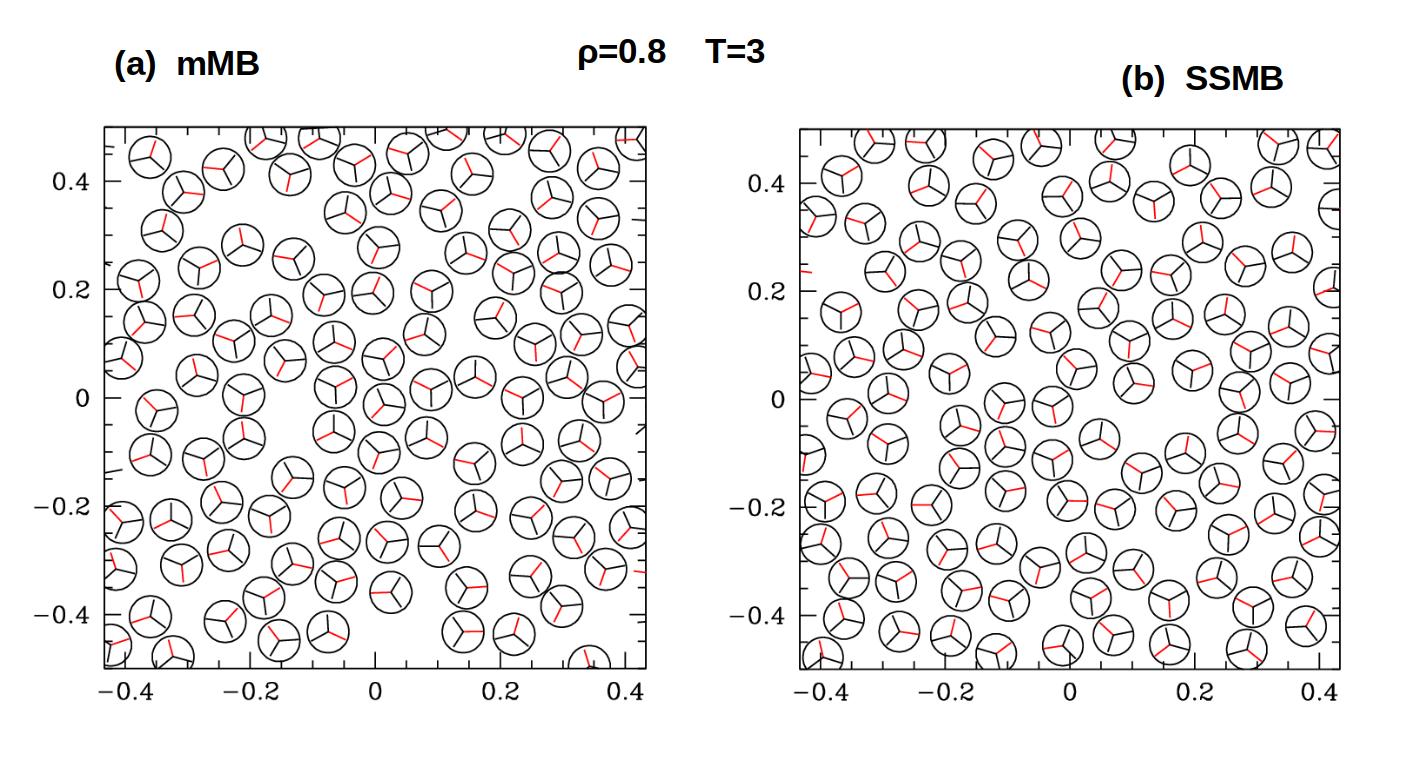}

\caption{Snapshot comparison between the mMB and the SSMB models, for number
of particles N=121. The red arm is assigned such that in the starting
configuration, all the red arms point along the y-axis.}

\label{FigSNAP1} 
\end{figure}

Fig.\ref{FigSNAP1} shows a comparison of particle configurations
(NVT ensemble simulations) between the mMB and SSMB models, for density
$\rho=0.8$ and temperature $T=3$. Both configurations appear quite
similar, particularly in terms of Hbond ordering. It seems, however,
that the SSMB model exhibits visually more homogeneous Hbond ordered
particles, which may explain why its structure factor does not exhibit
the sharp $k=0$ raise observed for both the MB and mMB models.

The origin of this homogeneity may come from the long ranged Yukawa
interaction, while the sharper Gaussian interaction cutoff of the
MB and mMB models may favour decorrelation between Hbonded clusters,
hence more disordered general distribution.

The influence of this type of disorder may be interesting to study
in the context of the presence of solutes and aqueous mixtures in
general.

\subsection{Thermodynamical anomalies of the SSMB model}

\begin{figure}[H]
\includegraphics[scale=0.3]{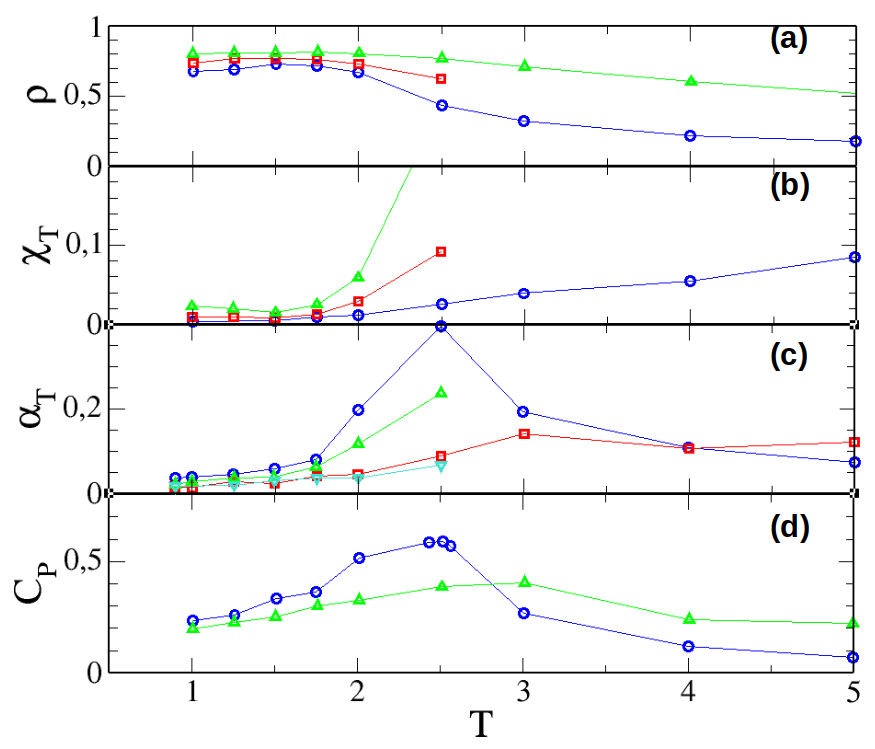}

\caption{Thermodynamic anomalies of the SSMB water model. (a) density maximum;
(b) isothermal compressibility minimum; (c) volume expansivity minimum
(see text); (d) isobaric heat capacity maximum. The blue circle is
for pressure $P=$1, the red square for $P=$2 and the up triangle
for $P=5$. Data for $P=10$ are shown in cyan for $\alpha_{T}$ in
(c).}

\label{FigANOM} 
\end{figure}

The thermodynamic anomalies, as obtained from the SSMB model through
Eqs.(\ref{eq:fluctuations}) are shown in Fig.(\ref{FigANOM}). We
use the following standard reduced units as: the reduced pressure
is $P=p\sigma^{2}/\epsilon_{LJ}$, the reduced compressibility is
simply given by $\chi_{T}=S(k=0)$, the reduced thermal expansivity
is $\alpha_{T}=\alpha_{T}k_{B}T$ and the reduced heat capacity $C_{P}=C_{P}/k_{B}$
where $k_{\ensuremath{B}}$ is the Boltzmann constant. 

The density maximum is found in (a) around $T\approx1.6$ for pressure
$P\approx5$ and $T\approx1.2$ for $P=2$. It is not well defined
for $P=5$. The isothermal compressibility in (b) has a clear minimum
for $P=2$ and $P=5$, around $T\approx1.5$. The isobaric heat capacity
maximum $C_{P}$ in (d) has also a well defined maximum whose temperature
dependence increases with increasing pressures.

The volume expansivity in (c) does not have a clear minimum, as expected
in the low temperature region. A calculation for higher pressure of
$P=10$ (cyan curve) does not show any more evidence. In addition,
it does not tend to become negative at low temperatures as for the
MB model and real water. This is an indication that volume fluctuations
and entropy fluctuation remain correlated as $T$ is decreased, instead
of becoming anti-correlated as in the case of real water. It is difficult
to say if this is a flaw of the SSMB model, since it is not obvious
how dimensional reduction, while keeping water-like spatial structure,
should affect volume and entropy correlations. This point remains
to be explored further. Since the SSMB model does not have the same
high density fluctuation at $k=0$ as the MB and mMB models, the absence
of expansivity anomaly could be related this property. Indeed, as
commented above, the higher spatial homogeneity of the SSMB model,
compared with that of the MB model, would imply a lower entropy, hence
a higher coupling between volume and entropy fluctuations, at the
origin of the absence of anomaly in the expansivity. In any case,
it is interesting that this SSMB model should have only some of the
water anomalies, but not all. It indicates that dimensional aspects
could play an important role. If this is true, then it would imply
that, for real water, these anomalies are not necessarily interrelated
and could not depend solely on Hbond properties.

\subsection{Low temperature phases}

\begin{figure}[H]
\includegraphics[scale=0.3]{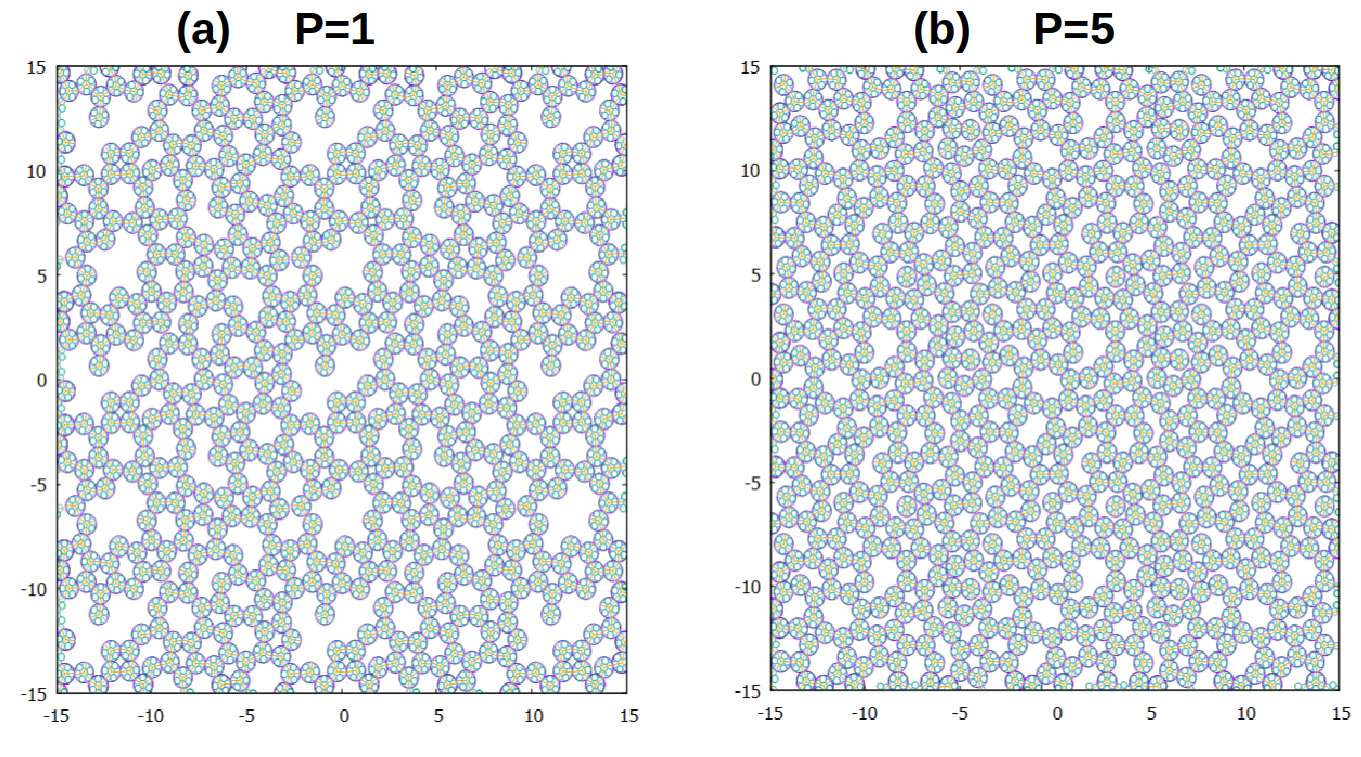}

\caption{Low temperature amorphous ice phases of the SSMB water model, for
T=0.9 and for 2 different pressures.}

\label{FigAMORPH} 
\end{figure}

Fig.\ref{FigAMORPH} shows the type of the low temperature $T=0.9$
amorphous ice that are found for the SSMB. Larges patches of Hbond
connected particles are observed at both low and high pressure. But
no perfect hexagonal crystal is found. We recall that the MB model
crystallises in a perfect hexagonal crystal\cite{mbs}. It should
be reminded that hard discs crystalline phases present alignment distortions,
which are generated by the density fluctuations in 2D being larger
than in 3D, thus avoiding perfect crystals in 2 dimensions, which
is the reason why the solid-liquid melting transition is not first
order, but second order with an intermediate Kosterlitz-Thouless phase\cite{NelsonHalperin,Frolich}.
The dimensionality plays an important role in phase transitions, and
it can be rigorously shown than continuous transitions (unlike that
in the Ising model) such as the XY model and Heinsenberg model, are
impossible in 2D\cite{Mermin-Wagner}. The liquid-solid transition
in 2D is itself controversial \cite{strandburg}.

In this context, by allowing core contact, the MBSS model meets some
of these problems common to 2D spherical core interactions. It is
then not so surprising that this model does not exhibit a perfect
hexagonal crystal. The snapshots of Fig.\ref{FigAMORPH} show several
clusters of hydrogen bonded pentamer, hexamers, up to disorted octamers.
The denser phase for $P=5$ equally shows flattened high order $n$-mers.
It does not seem surprising that a realistic 2D water model may not
have a perfect crystalline phase.

\section{Discussion and Conclusion}

We have introduced a site-site two-dimensional water model, named
SSMB model, as an alternative to the usual MB model. We have showed
that this new model transfers the dual Hbond / core contact paradigm
of water, from real space into reciprocal space, in agreement with
the real water model. This is an important feature to note, since
it is generally advertised that water has this dual order, such as
for instance in the Franks water model\cite{Franks} , implicitly
implying an order in real space. The specificity of the water structure
has been debated several times\cite{small-k-2} , and more recently
in the context of the water second critical point\cite{2ndCP}, which
would be related to a phase separation between disordered and Hbond
ordered 2 liquids, albeit in metastable conditions. It is not obvious
that such issues would be relevant to two-dimensional water models.
The MB model has been very successful in allowing to visually illustrate
in 2 dimensions many of the water anomalies and specificity in the
context of mixing. Although the present work is specifically focused
on highlighting the structural aspects of the 2D water models, it
would be interesting to examine a site-site version of the MB models
extended to various solutes, such as alcohols and ionic species, part
of the project prospective which is being currently pursued.

\section*{Acknowledgments}

It is a pleasure to participate to this special issue for Miroslav
Holovko. The authors (AP and TU) thank the partenariat Hubert Curien
(PHC) from Campus France for financial support under the bilateral
PROTEUS PHC project 44072XC. TU thanks for the financial support of
the Slovenian Research Agency through Grant P1-0201 as well as to
projects N1-0186, L2-3161 and J4-4562 as well as National Institutes
for Health RM1 award RM1GM135136. TB and MB thank the Laboratoire
de Physique Théorique de la Matière Condensée for hosting during their
internship.

\end{document}